\documentclass[12pt,epsf,epsfig,aps,floatfix,preprint,nofootinbib]{revtex4}
\usepackage{graphics}
\usepackage{graphicx}
\usepackage[english]{babel}
\usepackage{fullpage}
\usepackage{amssymb}
\usepackage[final]{pdfpages}
\usepackage{amsmath}
\usepackage{verbatim}
\usepackage{pstricks}
\usepackage{slashed}

\newcommand{\nc}{\newcommand}
\nc{\beq}{\begin{equation}}
\nc{\eeq}{\end{equation}}
\nc{\barray}{\begin{eqnarray}}
\nc{\earray}{\end{eqnarray}}
\nc{\barrayn}{\begin{eqnarray*}}
\nc{\earrayn}{\end{eqnarray*}}
\nc{\bcenter}{\begin{center}}
\nc{\ecenter}{\end{center}}
\nc{\ket}[1]{| #1 \rangle}
\nc{\bra}[1]{\langle #1 |}
\nc{\0}{\ket{0}}
\nc{\mc}{\mathcal}
\nc{\er}[1]{(\ref{eq:#1})}
\nc{\onehalf}{\frac{1}{2}}
\nc{\partialbar}{\bar{\partial}}
\nc{\psit}{\widetilde{\psi}}
\nc{\Tr}{\mbox{Tr}}
\nc{\ev}{\;\mathrm{eV}}
\nc{\mev}{\;\mathrm{MeV}}
\nc{\gev}{\;\mathrm{GeV}}

\def\chii0{\chi_i^0}
\def\chij0{\chi_j^0}

\newcommand{\bi}{\begin{itemize}}
\newcommand{\ei}{\end{itemize}}

\newcommand{\gsim}{\lower.7ex\hbox{$\;\stackrel{\textstyle>}{\sim}\;$}}
\newcommand{\lsim}{\lower.7ex\hbox{$\;\stackrel{\textstyle<}{\sim}\;$}}
\newcommand{\vrel}{v_{\text{rel}}}
\newcommand{\aem}{\alpha_{\text{em}}}
\newcommand{\mx}{m_{X}}
\newcommand{\mpl}{m_{\text{Pl}}}

\newcommand{\sigmaanave}{\left<\sigma_{\text{an}}\vrel\right>}

\newcommand{\xs}{x_{\text{s}}}
\newcommand{\xx}{x_{X}}

\newcommand{\vx}{v_X}

\begin{document}

\setlength{\baselineskip}{0.22in}

\begin{flushright}MCTP-10-52 \\
\end{flushright}

\vspace{0.2cm}

\title{Turning off the Lights: How Dark is Dark Matter?}

\author{
Samuel D. McDermott, Hai-Bo Yu,
Kathryn M. Zurek
}

\vspace*{0.2cm}

\affiliation{
Michigan Center for Theoretical Physics, Department of Physics, University of Michigan, Ann Arbor, MI 48109
}

\date{\today}

\begin{abstract}
\noindent

We consider current observational constraints on the electromagnetic charge of dark matter.  The velocity dependence of the scattering cross-section through the photon gives rise to qualitatively different constraints than standard dark matter scattering through massive force carriers.  In particular, recombination epoch observations of dark matter density perturbations require that $\epsilon$, the ratio of the dark matter to electronic charge, is less than $10^{-6}$ for $m_X = 1 \mbox{ GeV}$, rising to $\epsilon < 10^{-4}$ for $m_X = 10 \mbox{ TeV}$.  Though naively one would expect that dark matter carrying a charge well below this constraint could still give rise to large scattering in current direct detection experiments, we show that charged dark matter particles that could be detected with upcoming experiments are expected to be evacuated from the Galactic disk by the Galactic magnetic fields and supernova shock waves, and hence will not give rise to a signal.  Thus dark matter with a small charge is likely not a source of a signal in current or upcoming dark matter direct detection experiments.

\end{abstract}

\maketitle

\section{Introduction}

The nature of the dark matter (DM) remains a mystery.  For DM in the MeV to TeV range, a wide variety of probes constrain the DM to be a Weakly Interacting Massive Particle (WIMP) which interacts with ordinary matter through suppressed couplings.  These probes include direct detection of DM through nuclear recoils in underground detectors as well as indirect detection through DM annihilation to SM states in the sun (to neutrinos), in the Galactic center (to photons), and in the Galactic neighborhood (to charged particles).  There are also significant constraints on DM couplings to ordinary matter through production and escape as missing energy at colliders.  For a review, see \cite{review}.

Many of the most popular DM candidates naturally meet these stringent requirements.  The neutralino from supersymmetry, for example, carries no electric charge and can interact only sub-weakly, via the Higgs or through small couplings to the $Z$ boson, evading the most stringent constraints from LEP, Tevatron, and direct detection experiments such as CDMS \cite{CDMS} and XENON10 \cite{XENON}.  Its thermal annihilation cross-section is below the bounds for indirect detection through neutrinos, photons, or charged cosmic rays.  While WIMP DM has escaped direct and indirect detection thus far, it may be within reach. Direct detection experiments are scaling up, the reach of the LHC will begin to encompass weak scale DM candidates soon, and Fermi will continue to constrain DM annihilation in dwarf galaxies, the Galactic center, and in the halo.

At the same time, it is desirable to take as model-independent an approach as possible when constraining the nature of the DM.  While in most popular models the DM carries no electromagnetic charge, periodically the notion of CHArged Massive Particle (a CHAMP) has reappeared in the literature \cite{champ1,champ2,champ3,champ4,champ5}.  In some of the earliest discussions of CHAMPs, the DM carried a full unit of charge, but it was realized that this runs into a wide range of very stringent constraints from searches for heavy hydrogen to direct detection in underground labs.  Some of these constraints may not apply if the CHAMPs are expelled from the disk via shock waves from supernova remnants and screened from re-entry by the Galactic magnetic fields \cite{Chuzhoy:2008zy}.   More recently, the possibility that DM carries a fractional or epsilon-charge has been considered and constrained via the CMB acoustic peaks \cite{Dubovsky:2003yn}. Radio observations also constrain the electronic charge of the dark matter~\cite{Gardner:2009et}. In addition, the notion that the DM carries a ``dark charge'' has recently been considered \cite{Feng:2008mu,Ackerman:2008gi,Feng:2009mn,Ibarra:2008kn,Kaloper:2009nc,Dai:2009hx}. In these latter models the DM does not couple to the photon, but to a massless gauge force in the hidden sector.

In light of the current understanding of structure formation and cosmological history, we determine how large the DM charge can be while remaining consistent with current constraints.  We also consider direct detection signals from epsilon-charged DM, and determine whether it is possible to give rise to the signals in DAMA \cite{dama} and CoGeNT \cite{cogent} as discussed recently in \cite{foot}.  Because we are answering a general question about the coupling of DM to the photon, we leave our discussion of models to a minimum.  We note that the discussion encompassed by this paper does bring to light a number of constraints that strongly disfavor some recent models in the literature.   We comment on these models below where relevant.  DM may also have a magnetic or electric dipole; this has been thoroughly considered recently \cite{Sigurdson}, and we do not discuss it here.

The outline of this paper is as follows.  We begin with a brief discussion of models and the implications of this study for the viability of these models.  We then review the relic density calculation before turning to constraints.  We discuss halo shape constraints and the bound from scattering at recombination times.  We discuss direct detection of charged particles in light of the signals from CoGeNT and DAMA, and the implications of the bounds discussed here for these experiments and models designed to fit them.  Finally, we conclude.

\section{Models and General Considerations}

Since DM that carries an electric charge must conserve $U(1)_{\rm EM}$, it must be a Dirac particle.  There are a number of models in the literature where the DM carries a fractional or epsilon-charge.  If a dark photon is massive and kinetically mixes with the photon, an epsilon-charge arises in Stueckelberg models \cite{Feldman} on account of the unique form of Stueckelberg mass term.
If, on the other hand, the dark photon is massless, kinetic mixing between the dark and visible photons induces an electric charge for the DM (or equivalently, a dark charge for visible states) \cite{Holdom:1985ag}.  This mechanism is utilized for example in the Mirror Charged DM model proposed by \cite{foot} to generate the signals in CoGeNT and DAMA.  We will see that the constraints we discuss here strongly disfavor such a model as the explanation for these signals.  In either case, we denote the charge of the DM as $\epsilon e$.

When determining the constraints on the DM charge, the essential features will be the irreducible coupling to the photon (and charged SM particles), and, more importantly, the velocity dependence of the scattering cross-section.  For example, the Rutherford Scattering cross-section of DM off DM through a photon is
\begin{equation}
\frac{\mathrm{d} \sigma_{XX}}{\mathrm{d} \Omega_*} = \frac{\aem^2 \epsilon^4}{m_X^2 v_{\rm rel}^4 \sin^4(\theta_*/2)},
\label{rutherford}
 \end{equation}
where $m_X$ is the DM mass, $v_{rel}$ is the DM relative velocity, and $\theta_*$ is the scattering angle in the center-of-mass frame.  Likewise, the scattering cross-section of DM off baryon is
\begin{equation}
\frac{\mathrm{d} \sigma_{Xb}}{\mathrm{d} \Omega_*} = \frac{\aem^2 \epsilon^2}{4 \mu_b^2 v_{\rm rel}^4 \sin^4(\theta_*/2)},
\label{Xb}
\end{equation}
where $\mu_b$ is the DM-baryon reduced mass.

The important point phenomenologically is the very large enhancement in the scattering cross-section at low velocity, giving a hint for where to look for strong constraints on DM charge.  Galactic constraints, where the DM has been heated through collapse and virialization, as we will see, tend to give weak constraints.  In contrast, the tightest constraints come primordially, before collapse and heating occur, when the DM is highly non-relativisitic.  In particular we will find that constraints on DM coupled to baryons at the time of recombination and DM coupled to baryons in protohalos can be very important, and this constraint will eliminate models whose charges are larger than about $10^{-6}$, dependent on the mass of the DM.  This constraint eliminates a broad class of models.

On the other hand, this constraint does not eliminate DM with epsilon-charges that can give rise to a signal in direct detection experiments.  We will find, however, that in the region where the DM could give rise to a signal in direct detection, one expects the DM to have been evacuated from the disk via supernova shock waves, and its re-entry to have been prevented by Galactic magnetic fields.  Therefore, although direct detection experiments are extremely sensitive to small charges, we will find that charged DM, such as suggested in \cite{foot}, could not plausibly give rise to a signal in a direct detection experiment.  We now go through these constraints in detail.

\section{Relic Density Constraints}
We begin by discussing the constraints from the relic density.
If the DM is non-thermally produced, its relic density depends on the production mechanism (for example, if the DM particle is produced via the decay of a mother particle, its relic density depends on the number density of the mother particle). In this scenario, constraints from the current relic abundance are highly model dependent.  On the other hand, in the case of thermal relics, the DM density is simply determined by the thermally averaged annihilation cross section.  As we will discuss explicitly, a charged DM consistent with all cosmological constraints must be non-thermally produced, unless it has additional interactions.  We now review the relic density considerations.

The DM can annihilate to photon pairs and to charged fermion pairs through the photon. In general, the charged DM can also carry other SM or hidden sector quantum numbers and annihilate through these channels as well. In our analysis we will not specify these additional interactions in detail; instead, we maintain a less model-dependent view. We assume DM is in thermal equilibrium in the early universe and require the irreducible annihilation processes not overly deplete DM. By considering only the annihilation channels induced by the electromagnetic charge of the DM, we can derive upper bounds on the charge $\epsilon$ for a given mass $\mx$.

The annihilation cross sections of $X\bar{X}\rightarrow\gamma\gamma$ and $f\bar{f}$ at tree level are given by
\begin{equation}
\left(\sigma_{\text{an}}\vrel\right)_{\gamma\gamma}=\frac{\pi\aem^2\epsilon^4}{m^2_X}
\end{equation}
and
\begin{equation} \label{sigmaxxff}
\left(\sigma_{\text{an}}\vrel\right)_{f\bar{f}}=\frac{\pi\aem^2\epsilon^2}{m^2_X}q^2_fN_c\sqrt{1-\frac{m^2_f}{m^2_X}}\left(1+\frac{m^2_f}{2m^2_X}\right),
\end{equation}
respectively, where $q_f$ is the charge of the SM fermion in units of electron charge and $N_c$ is the color multiplicity of the fermion. The total annihilation cross section of the DM particle at tree level is $\left(\sigma_{\text{an}}\vrel\right)_{\text{tot}}=\left(\sigma_{\text{an}}\vrel\right)_{\gamma\gamma}+\sum_{f} \left(\sigma_{\text{an}}\vrel\right)_{f\bar{f}}$.

 The tree level annihilation cross section is enhanced by the Sommerfeld effect in the low velocity dispersion \cite{Sommerfeld:1931,Baer:1998pg,Hisano:2002fk,sommerfeld}. DM freeze out with Sommerfeld enhancement has been discussed in~\cite{freezeout1,freezeout2}. Since the mediator of the Sommerfeld enhancement is the standard model photon with zero mass, this enhancement never saturates. The enhancement factor for the tree level S-wave annihilation cross section is given by
\begin{equation}
S=\frac{(\aem\epsilon^2\pi)/v}{1-e^{-(\aem\epsilon^2\pi)/v}},
\end{equation}
where $v=\vrel/2$ is the DM velocity in the center of mass frame. The thermally averaged total annihilation cross section including the Sommerfeld enhancement is given by
\begin{equation}
\sigmaanave_{\text{tot}}=\left(\sigma_{\text{an}}\vrel\right)_{\text{tot}}\frac{\xx^{3/2}}{2\sqrt{\pi}}\int^\infty_{0} S \vrel^2 e^{-\xx\vrel^2/4}d\vrel,
\end{equation}
where we assume a Maxwell-Boltzmann distribution for DM particle and $\xx \equiv \mx/T_{X}$ with $T_X$ as the DM temperature. The Sommerfeld enhanced annihilation tends to deplete DM particles with low velocity, which may distort the thermal distribution of the DM after kinetic decoupling. However, as we will show in the next section, the charged DM can couple to the thermal bath even during the recombination epoch, and therefore the Maxwell-Boltzmann distribution is a good approximation.

Following the standard procedure to calculate the abundance of a thermal relic~\cite{Gondolo:1990dk,Kolb:1990vq}, freeze out occurs when
\begin{eqnarray}
x_f&\approx&\ln\xi-\frac{1}{2}\ln\left(\ln\xi\right)\\
\xi&=&0.038\kappa(2+\kappa)m_{\text{pl}}\mx(g/\sqrt{g_*})(\sigma_{\text{an}}\vrel)_{\text{tot}},
\end{eqnarray}
where $x=\mx/T$, $T$ is the temperature of the thermal bath, and $g$ is the number of degrees of freedom of the DM particle; we take $g=4$, for a Dirac particle. The value of $\kappa$ is chosen to match the numerical solution; we set $\kappa=1$.

The present number density of the DM is the solution of the Boltzmann equation, which can be written as
\begin{eqnarray}
\frac{1}{Y(\xs)} &=& \frac{1}{Y(x_f)}
+ \sqrt{\frac{\pi}{45}} \mpl \, m_X  \int_{x_f}^{x_s}
\frac{(g_{*s}/\sqrt{g_*})\sigmaanave_{\text{tot}}}{x^2} dx ,
\label{relic}
\end{eqnarray}
where $Y=n_{X}/s$ with $s$ the entropy density, and $x_s=T_{\rm s}/\mx$ with $T_{\rm s}=1~{\rm eV}$, where we stop the integration. Here we assume $X$ and $\bar{X}$ have the same number density, $n_X=n_{\bar{X}}$, and the total number density of the DM is their sum, $n_{X}$+$n_{\bar{X}}$. Before kinetic decoupling, the DM temperature is the same as the thermal bath temperature and drops as $a^{-1}$, where $a$ is the scale factor. After kinetic decoupling, the DM temperature drops as $a^{-2}$. Therefore, the enhancement factor scales as $S\sim x^{-1}$ and $S\sim x^{-2}$ with respect to $x$ before and after kinetic decoupling, respectively~\cite{freezeout1}. Since DM is tightly coupled to baryons through the massless photon until after the recombination epoch, $S\sim x^{-1}$ over the entire range of integration of Eq.~\eqref{relic}.  As we will show in Section IV, the elastic scattering rate rises as the temperature drops so that the DM may still be coupled to the thermal bath when the temperature is below $1~{\rm eV}$, even if $\epsilon$ is chosen to satisfy the relic density bound.

\section{Structure Formation and CMB Constraints}

\subsection{Decoupling at the Recombination Epoch}

\begin{figure}
  \centering
    \includegraphics[width=0.8\textwidth]{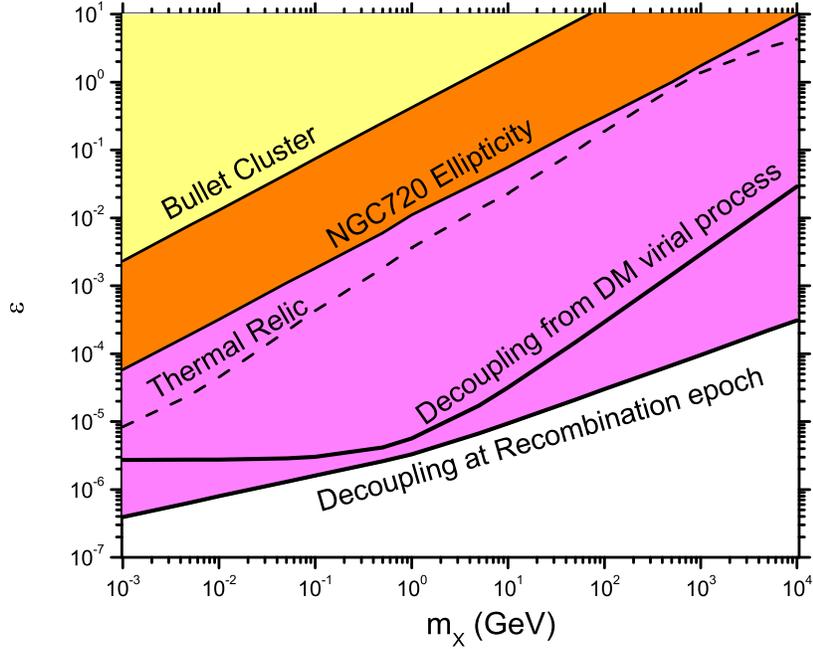}
     \caption{Constraints from various sources, from top to bottom:
(i) Scattering in the bullet cluster and NGC720, (ii) DM as a charged thermal relic, and (iii) DM virial processes, and (iv) recombination epoch.}
\label{master}
\end{figure}

Charged DM particles interact with the Standard Model via a small coupling through the photon, so that Coulomb scattering can couple the charged DM to the baryon-photon plasma tightly even at low temperature. If the DM is still in kinetic equilibrium with the baryon-photon plasma during recombination, the DM density fluctuations can be washed out due to the radiation pressure and the photon diffusion (Silk damping~\cite{Silk:1967kq}). The baryon acoustic peak structure will also be directly altered through the coupling.
The effects of millicharged particles on CMB acoustic peaks have been discussed in Refs.~\cite{Dubovsky:2003yn,Burrage:2009yz}.  It was found that if the millicharged particles couple to the baryon-photon plasma tightly during the recombination epoch they behave like baryons, and CMB observations put an upper limit on their abundance $\Omega_{\rm mcp}h^2<0.007~(95\%)$~\cite{Dubovsky:2003yn}. 
Here we assume that the DM is made of epsilon-charged particles and derive the relaxation time scale for the DM to reach kinetic equilibrium with baryons. To avoid damping effects on DM density fluctuations and CMB anisotropy constraints, we require that DM have {\em completely} decoupled from the photon-baryon plasma at the recombination epoch, and derive a bound on $\epsilon$ for a given DM mass. 

We consider a DM particle that has momentum $p_X$ in its comoving frame. After each scattering event the magnitude of the DM momentum changes by an amount $\delta p_X$. The momentum transfer rate is thus given by
\begin{equation} \label{momentumtransfer}
\Gamma_p = \frac{\mathrm{d}\langle \delta p^2_X \rangle/\mathrm{dt}}{\left<p^2_X\right>}.
\end{equation}
The thermally averaged momentum transfer per unit time is
\begin{equation} \label{psquaredot}
\mathrm{d}\langle \delta p^2_X \rangle/\mathrm{dt}=\sum_{b=e,p}n_b\int d^3v_B d^3v_X f(v_B)f(v_X)d\Omega_*\frac{d\sigma_{Xb}}{d\Omega_*}\vrel\delta p^2_X,
\end{equation}
where $d\sigma_{Xb}/d\Omega_*$ is given by Eq.~\eqref{Xb}, $n_b$ is the number density of the baryon, and  $\delta p^2_X$ is the momentum transfer after one collision:
\begin{equation}
\delta p^2_X=2\mu^2_b\vrel^2(1-\cos\theta_*).
\end{equation}
Note that this quantity is reference frame independent.  
The thermally averaged momentum squared of the DM particle in its comoving frame is
\begin{equation} \label{momentumaverage}
\langle p^2_X \rangle=\int d^3v_Xf(v_X) (\mx v_X)^2=\frac{3}{2}m_X^2 v_0^2=3m_XT
\end{equation}
for a DM particle in a thermal Maxwell distribution.
To evaluate the thermal average for $\vrel^2$, we derive a general formula. For a given function of $g(\vrel)$, we have
\begin{equation}
\int d^3v_ad^3v_bf(v_a)f(v_b)g(\vrel)=\int d\vrel \vrel^2 \frac{4}{\sqrt{\pi}}\frac{1}{(v^2_{0a}+v^2_{0b})^{\frac{3}{2}}}e^{-\frac{\vrel^2}{v^2_{0b}}+\frac{\vrel^2v^2_{0a}}{(v^2_{0a}+v^2_{0b})v^2_{0b}}}g(\vrel),
\end{equation}
where we assume $f(v_{a,b})$ are Maxwellian distributions and $v_{0a,b}$ are the most probable velocities for the $a$ and $b$ particles, respectively. By using this general formula, we have
\begin{equation}
\mathrm{d}\langle \delta p^2_X \rangle/\mathrm{dt}=-\sum_{b=e,p}\frac{8\sqrt{2\pi}n_b\aem^2\epsilon^2}{\left(\frac{T}{\mx}+\frac{T}{m_b}\right)^{\frac{1}{2}}}\ln\left(\theta^{\text{min}}_*/2\right),
\end{equation}
where $\theta^{\text{min}}_*$ is the cutoff in the $\Omega_*$ integral.  Its value is set by the maximum impact parameter due to Debye screening effects in the plasma. This maximum impact parameter is related to the minimum scattering angle through
\begin{equation}
b_{\text{max}}=\frac{\aem\epsilon}{\left<\mu_b\vrel^2\right>}\cot(\theta^{\text{min}}_*/2),
\end{equation}
where $\left<\mu_b\vrel^2\right>=3T$ and $\cot(\theta^{\text{min}_*}/2)\simeq2/\theta^{\text{min}}_*$ for small $\theta^{\text{min}}_*$. The impact parameter of the scattering must not be larger than the Debye screening length, so we have $b \leq b_{\text{max}} =\lambda_D$. Thus
\begin{equation} \label{thetamin}
\theta^{\text{min}}_*\simeq\frac{2\epsilon\aem}{3T\lambda_D},
\end{equation}
where $\lambda_D=\sqrt{T/(4\aem\pi n_e)}$ is the Debye length for the baryon plasma. Using Eq. \eqref{momentumaverage}, we then have a momentum transfer rate of:
\begin{equation} \label{CMBconstraint}
\Gamma_p=\sum_{B=e,p}\frac{8\sqrt{2\pi}n_b\aem^2\epsilon^2\mu^{\frac{1}{2}}_b}{3\mx T^{\frac{3}{2}}}\ln\left[\frac{3T\lambda_D}{\epsilon \aem}\right].
\end{equation}
If the DM is tightly coupled to the baryon-photon plasma during recombination, DM density fluctuations will be damped. CMB observations also place strong constraints on the total abundance of charged particles in the tightly-coupled regime. Here we require that the relaxation time of the momentum transfer rate is larger than the Hubble time at the recombination epoch; {\em i.e.},
\begin{eqnarray}
\Gamma^{-1}_p(T_R)>t_R,
\end{eqnarray}
where $t_R\simeq 3.8\times10^{5}~\text{years}$~\cite{nasa}, and $T_R\simeq0.26~{\rm eV}$ is the temperature at the recombination epoch. We take the baryon number density at recombination $n_e=n_p=\Omega_b\rho_c a^{-3}_R/m_p$, where $\Omega_b\simeq0.023h^{-2},~\rho_c=8.0992h^{2}\times10^{-47}~{\rm GeV^4}$, $a_R\simeq1/1091$, and $h\simeq0.71$~\cite{nasa}. This constraint is plotted in Figure \ref{master}.  In the above analysis, we have implicitly assumed that photons couple to baryons efficiently during the recombination epoch despite the presence of charged DM.  We checked that electron-photon Compton scattering can keep baryons and  DM  particles in the kinetic equilibrium with photons. This is because the DM density is not far from the baryon density. The same Compton drag force also suppresses the growth of the DM density perturbations.

Here we ignore the process of DM-photon Compton scattering. Since the cross section $\sigma_{X\gamma}=8\pi\aem^2\epsilon^4/(\mx^2)$ is proportional to $\epsilon^4$ and the momentum transfer rate through this process is also highly suppressed kinematically at low temperature, we expect that the bound derived from the DM-photon decoupling is weak. As shown in Ref.~\cite{Boehm:2001hm}, CMB anisotropies and matter power spectrum requires $\sigma_{X\gamma}/\mx<10^{-32}~{\rm cm^2~GeV^{-1}}$. We can translate this limit to a bound on $\epsilon$ as $\epsilon<0.49~(\mx/(1~{\rm GeV}))^{3/4}$, which is much weaker than the bound derived from the DM-baryon decoupling.

\subsection{Effect on the Dark Matter Virialization}

After recombination, radiation damping suppression is absent, but the efficient energy transfer between baryons and charged DM particles will modify the virialization process of the DM. Since baryons decouple from the thermal bath much later than DM particles, baryons are hotter than DM particles at redshift $z\sim30$ when protohalos start to form. If there is a tight coupling between DM particles and baryons at this epoch, baryons will transfer energy to DM particles and heat them up. We can derive a bound on $\epsilon$ by requiring the energy transfer time be longer than the DM virialization time.

Eq.~\eqref{CMBconstraint} is no longer valid for charged particles with slow motion in a neutral medium. At these late epochs, although it appears that the Born approximation condition $\epsilon^2\aem/\vrel<1$ may still be satisfied due to the smallness of $\epsilon$, the charged DM particle typically has a wavelength larger than the Bohr radius of the hydrogen atom, and one must take into account the screening effect.
This effect is analogous to the energy loss of a slow-moving ion in the neutral medium, a result first derived by Lindhard and Scharff~\cite{Lindhard:1961zz}. Lindhard's approximation is valid when the impact parameter is larger than the Bohr radius.  When the protohalo forms at redshifts $z \sim 30$, the DM velocity dispersion is $\mathcal{O}(10^{-8}c)$. Its de Broglie wavelength is much larger than the Bohr radius under these conditions, and we expect that Lindhard's formula applies. We calculate the energy exchange of the charged DM in the hydrogen medium using Lindhard's formula,
\begin{eqnarray}
\frac{dE_X}{d\ell}=\frac{n_H}{m_e}\left[\frac{\pi^2 \epsilon}{2.7183}\frac{m_X}{m_X+m_H}\right],
\end{eqnarray}
where $n_H$ is the hydrogen number density, and we ignore the negligible effects of electron recoil~\cite{Lindhard:1961zz} and other elements. The relaxation time scale is estimated as
\begin{eqnarray}
\tau_X\simeq\left<E_X\right> \left<\frac{1}{\vrel} \frac{d\ell}{dE_X}\right>=\frac{3\times2.7183m_e(m_H+m_X)}{4\sqrt{2}\pi^{3/2}n_H\epsilon}\sqrt{\frac{T_H}{m_H}+\frac{T_X}{m_X}},
\end{eqnarray}
where $E_X =m_X v_{\rm rel}^2/2$.  We take $\sqrt{T_H/m_H}\sim 10^{-6}c$, and $\sqrt{T_X/m_X}\sim10^{-8}c$ at $z\sim30$.

In the usual cold DM scenario, DM collapses and virializes at a redshift of $z\sim30$. In over-dense regions the density is about $178$ times larger than the average density at the same epoch~\cite{Kamionkowski:2008gj}, and the violent relaxation time scale is
\begin{eqnarray}
\tau_{\rm vir}\sim(G\rho_{\rm tot})^{-1/2},
\end{eqnarray}
where $\rho_{\rm tot}=\rho_X+\rho_{\bar{X}}$ and $\rho_{\rm tot}\sim 178\bar{\rho}_{\rm tot}=178\Omega_X\rho_c(1+z)^3$. Now we demand $\tau_X>\tau_{\rm vir}$ and obtain an upper bound on $\epsilon$,
\begin{eqnarray}
\epsilon < 2.9\times10^{-6}\left(\frac{m_H+m_X}{1~{\rm GeV}}\right),
\end{eqnarray}
which is shown in Fig. 1.

\section{Dark Matter Halo Constraints}

The strongest constraints on the coupling of DM to the photon come from scattering considerations rather than annihilations because of the large scattering cross-section enhancement at low velocities.  In the previous section, we explored the effects at high redshift from observations of universe at recombination temperatures.  The constraints are particularly strong in this regime. However, lower redshift observations can also be used to test the charged DM hypothesis.

\subsection{Elliptical galaxies}

If the DM self-interaction through Coulomb scattering is strong enough to create an ${\cal O}(1)$ change in the momentum of DM particles within the age of galaxies, it will isotropize the velocity dispersion and lead to more spherical halos. The collisions also cause heat conduction from the hot outer parts to the cooler inner parts of DM halos, giving rise to the formation of a core with a shallow density profile. These expectations have been confirmed by simulations in the hard sphere scattering limit~\cite{Dave:2000ar,Yoshida:2000bx,Moore:2000fp,Craig:2001xw,Kochanek:2000pi,Spergel:1999mh}. In addition, observations of  elliptical DM halos in clusters constrain self-interactions~\cite{MiraldaEscude:2000qt}, while observations of elliptical DM halos in galaxies provide the strongest constraints on self-interacting DM models~\cite{Feng:2009mn,Feng:2009hw,Ibe:2009mk}. In this paper, we will follow the analysis of Ref.~\cite{Feng:2009mn,Feng:2009hw,Ibe:2009mk} and use the ellipticity of NGC 720 to derive the upper bound of the electric charge of the DM.

To estimate how the ellipticity of NGC 720 may be used to constrain the charge of the DM, we calculate the relaxation time due to momentum transfer. We then assume the relaxation time scale is the same as the time scale for isotropizing the mass distribution of the DM halo. By using Eq. \eqref{rutherford}, the thermally-averaged momentum transfer rate inside the halo can be evaluated from as
\begin{eqnarray}
\Gamma_e=-16\pi\aem^2\epsilon^4\frac{\rho_{\rm tot}}{3\mx^3 v^2_0}\int d\vrel \vrel^2\sqrt{\frac{2}{\pi}}\frac{1}{v^3_0}e^{-\frac{\vrel^2}{2v^2_0}}\frac{1}{\vrel}\ln\left(\theta^{\rm min}_*/2\right),
\end{eqnarray}
where $\rho_{\rm tot}=\rho_{\rm X}+\rho_{\rm\bar{X}}$ is the total DM density of the halo, and the minimal scattering angle $\theta^{\rm min}_*$ is given by $\theta^{\rm min}_*\simeq 4\aem\epsilon/(3\mx v^2_0\lambda_D)$ with $\lambda_D=\mx^2v^2_0/(8\pi\epsilon^2\aem\rho_{\rm tot})$ as the Debye screening length of the DM halo.

The elliptical galaxy NGC 720 is well-studied~\cite{Buote:2002wd,Humphrey:2006rv}. In Ref.~\cite{Humphrey:2006rv}, X-ray isophotes were used to extract the ellipticity of the underlying matter distribution, and the DM halo of NGC 720 was found to be elliptical at 5 kpc and larger radii. At $5~{\rm kpc}$, the DM density is $\rho_{\rm tot}=4~{\rm GeV/cm^3}$, and the radial velocity dispersion $\overline{v^2_r}=v^2_0(r)/2\simeq(240~{\rm km/s})^2$~\cite{Feng:2009hw}.
To derive the constraints on $\epsilon$ for the given $\mx$ from the observed halo shapes, we require
\begin{eqnarray}
\Gamma^{-1}_e>10^{10}~{\rm years}.
\end{eqnarray}
That is, the average time scale to create ${\cal O}(1)$ change in the DM particle momentum must be greater than the galaxy's lifetime. This bound, weaker than the constraint derived from decoupling at the time of recombination, is depicted in Fig. \ref{master}.

\subsection{The Bullet Cluster}

In the Bullet Cluster system, a subcluster has collided with and moved through a larger cluster. These clusters have three major components that each behave very differently during the collision: stars, gas, and DM. The visible stars pass through without colliding, but the highly collisional X-ray gas slows down significantly. Gravitational lensing shows that the DM tracks the stars, which are effectively collisionless. These observations have been used to place stringent bounds on the self-interaction of the DM~\cite{Markevitch:2003at}. These bounds are derived through different considerations including the offset between the gas and DM, the high velocity of the subcluster, and the survival of the subcluster after the collision. It turns out that the survival of the subcluster puts the strongest bound on the self-interaction of DM~\cite{Markevitch:2003at}.

We follow the approach of Ref.~\cite{Markevitch:2003at} to derive bounds on the DM charge. The analysis of Ref.~\cite{Markevitch:2003at} is based on a hard sphere scattering cross section, but we relax this assumption. For Rutherford scattering, the subcluster experiences a net loss of DM particles if particles in both the main cluster and the subcluster have velocities larger than the escape velocity of the subcluster. We define the scattering angle $\theta$ to be measured in the rest frame of the subcluster, which implies $\theta=\theta_*/2$, where $\theta_*$ is the scattering angle in the frame of the center mass of two colliding particles.

The particle loss condition detailed above can be satisfied if $\sin\theta$ is in the following range:
\begin{equation}
\frac{v_{\rm esc}}{v_1}<\sin\theta <\sqrt{1-\frac{v^2_{\rm esc}}{v^2_1}},
\label{losscondition}
\end{equation}
where $v_1\sim4800~{\rm km/s}$ is the velocity of the main cluster incoming particles before the collision and $v_{\rm esc}\sim1200~{\rm km/s}$ is the escape velocity of the subcluster. We assume that the subcluster sees the main cluster with a surface number density $\Sigma_m\sim 0.3~{\rm g/cm^3}$~\cite{Markevitch:2003at}, and demand that the particle loss fraction $f$ be smaller than $30\%$, i.e.
\begin{equation}
f=\frac{\sum_m}{\mx}\int d\Omega_*\frac{d\sigma_{XX}}{d\Omega_*}=\frac{\sum_m}{\mx}\frac{4\pi\aem^2\epsilon^4}{\mx^2 v^2_1}\left(\frac{1}{v^2_{\text esc}}-\frac{1}{v^2_1-v^2_{\rm esc}}\right)<30\%.
\end{equation}
The Bullet Cluster bound is given in Fig. \ref{master}.

\section{Direct Detection of Charged Dark Matter}

\begin{figure}
  \centering
    \includegraphics[width=0.8\textwidth]{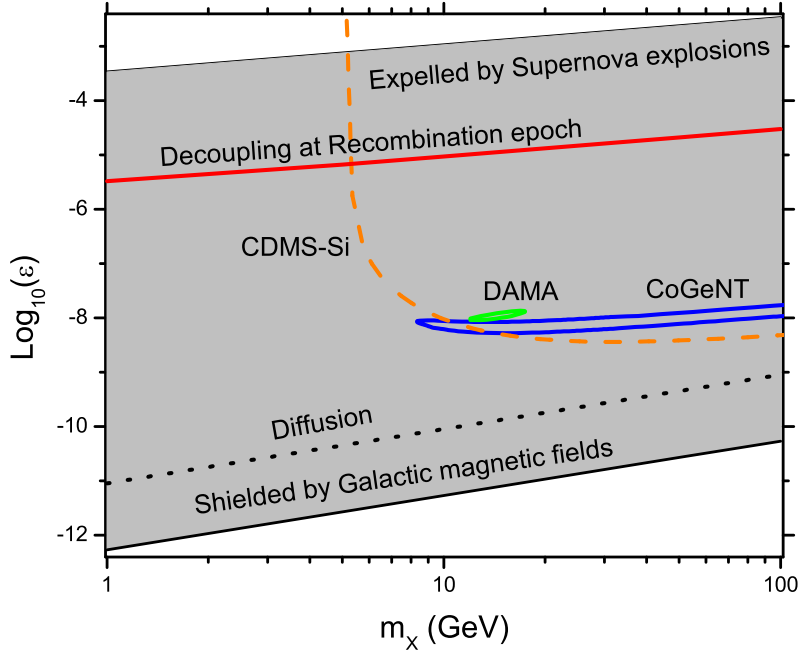}
     \caption{CoGeNT (blue), DAMA (green) allowed regions at 99\% C.L.  The CDMS-Si (yellow) line is included as a sample exclusion at $99\%$ C.L.
     In the gray area, the charged DM is evacuated from the Galactic disk. Also shown the bound from Recombination epoch (red). Below the dotted line (black), charged DM may diffuse to the disk.}
\label{fig2}
\end{figure}

Because of the large enhancement of the scattering cross-section at low velocity, even DM with a very small charge can give rise to a large scattering cross-section in direct detection experiments.  In \cite{foot}, it was found, for example, that a charge of $\epsilon\sim 10^{-9}$ was sufficient to give rise to the relatively large signals in CoGeNT and DAMA.  Thus, if correct, direct detection experiments have a potential to give rise to even tighter constraints on epsilon-charged DM with mass in the range $m_X \sim 10 \mbox{ GeV} - 1 \mbox{ TeV}$.  We find, however, that in the range of charges where DM could give rise to a signal in a direct detection experiment the DM will necessarily have been efficiently evaporated from the Disk, and thus one expects no signal. We begin by a review of the signal in direct detection experiments.

\subsection{Direct Detection Basics}

The rate for scattering is
\beq
\frac{dR}{dE_R} = N_T \frac{\rho_\chi }{m_{\chi}} \int_{|\vec{v}| > v_{min}} d^3v v f(\vec{v},\vec{v}_e)\frac{d\sigma}{d E_R},
\label{totalrate}
\eeq
where
$
v_{min}=\frac{\sqrt{2 m_N E_R}}{2\mu_N},
$
and $\mu_N$ is the reduced mass of the nucleus-DM system.
We take the velocity distribution $f(\vec{v},\vec{v}_e)$ to be a modified Boltzmann distribution
\beq
f(\vec{v},\vec{v}_e) \propto \left(e^{-(\vec{v}+\vec{v}_e)^2/v_0^2}-e^{-v_{esc}^2/v_0^2}\right) \Theta(v_{esc}^2-(\vec{v}+\vec{v}_e)^2),
\eeq
where explicit expressions for the velocity integrals from this distribution can be found in \cite{liam}.
The additional term is to allow for a smooth cut-off of the velocity distribution near the Galactic escape velocity $v_{esc}$.
The Earth's speed relative to the Galactic halo is $v_e = v_\odot + v_{orb} \cos \gamma \cos[\omega(t-t_0)]$ with $v_\odot = v_0 + 12 \mbox{ km/s}$, $v_{orb} = 30 \mbox{ km/s}$, $\cos \gamma = 0.51$, $t_0 = \mbox{ June 2nd}$ and $\omega = 2 \pi/\mbox{year}$.  We take as a standard case $v_0 = 220 \mbox{ km/s}$, and we fix $v_{esc} = 500 \mbox{ km/s}$ and the local DM density $0.3~{\rm GeV/cm^3}$.

A standard calculation relates the differential rate for scattering off nuclei to the scattering rate off a nucleus $\sigma_N$,
\beq
\frac{d\sigma}{d E_R} = \frac{m_N \sigma_N}{2 \mu_N^2 v^2}.
\label{eq:ratenorm}
\eeq
For the standard spin-independent case, this rate is related to a scattering off protons, $\sigma_p$, through
\beq
\sigma_N = \sigma_p \frac{\mu_N^2}{\mu_n^2} \frac{\left[f_p Z + f_n(A-Z)\right]^2}{f_p^2} F^2(E_R),
\eeq
where $\mu_n$ is the DM-nucleon reduced mass and $f_p$ and $f_n$ are the DM couplings to the neutron and proton.
We set $f_n = 0$ since the coupling is assumed to be through the photon.   We make use of a Helm form factor
$
F(E_R) = \frac{3 j_1(q r_0)}{(q r_0)} e^{-(qs)^2 {\rm fm}^2 /2},
$
where
$
r_0 = \left((1.2 A^{1/3})^2-5 s^2\right)^{1/2} \mbox{ fm},
$
with $s = 1$.
The scattering cross-section off nuclei through the photon is
\beq
\sigma_N = \frac{16 \pi \alpha^2 \epsilon^2 Z^2 \mu_N^2}{q^4},
\label{chargecrosssection}
\eeq
which is to be inserted in Eq.~\eqref{totalrate} to obtain the total rate as a function of energy.  As an example of the typical DM charge $\epsilon$ that can be probed with direct detection experiments, we show the constraints one obtains from the CDMS, DAMA and CoGeNT experiments in Fig.~(\ref{fig2}).  One can see that the viable region is well below the structure formation constraints labeled in the figure. Note that in Ref.~\cite{foot}, the mirror DM velocity dispersion depends on the particle mass, and is typically smaller than the rotation speed $v_0 = 220~{\rm km/s}$, so the allowed regions for both DAMA and CoGeNT as well as the excluded region for CDMS shift to larger DM mass (larger than $~20~{\rm GeV}$) compared to the fitting presented in Fig.~{\ref{fig2}}. Since the shift is more significant for the light target nuclei, the mirror DM model features a DAMA region which is not excluded by CDMS, as depicted in~\cite{foot}. We next discuss how DM with charges in this range will have been evacuated from the disk at the present day, eliminating any possible signal in a direct detection experiment.

\subsection{Evacuation of Charged DM from the Disk}

The charged DM interacts with the magnetic fields of the Galaxy in addition to baryons in the disk. Since the large-scale magnetic field in the Milky Way is mostly parallel to the plane of the Galactic disk, the charged DM particle in the halo may not be able to penetrate the disk if its gyroradius is smaller than the height of the disk. This magnetic shielding effect for the millicharged particle has been discussed in the Ref.~\cite{Chuzhoy:2008zy}. The gyroradius is given by
\begin{equation}
R_{\rm g}\simeq5.4\times10^{-11} ~ \mathrm{pc} \left(\frac{m_X}{1~{\rm GeV}} \right) \left(\frac{1}{\epsilon} \right) \left(\frac{v_X}{270 ~ \mathrm{km/s}} \right) \left(\frac{5~\mu \mathrm{G}}{B} \right)<H_{\rm d},
\end{equation}
where $H_{\rm d}\sim100~{\rm pc}$ is the typical height of the Galactic disk. So the Galactic magnetic field prevents charged DM from entering the disk if $\epsilon>5.4\times10^{-13}~(\mx/{\rm GeV})$.

It is possible that some quantity of charged DM can remain in the disk from the time when the disk formed. Subsequently, however, shock waves generated by supernova (SN) explosions can blow these particles out of the disk if the acceleration time scale ($\tau_{\rm acc}\simeq10^7~{\rm years}$) is shorter than the cooling time scale~\cite{Chuzhoy:2008zy}. The cooling time scale due to the scattering with electrons is given by
\begin{eqnarray}
\tau_{\rm cool}=\frac{\mx m_e\vx^3}{8\pi\aem^2\epsilon^2 n_{ e}}\left[\ln\left(\frac{\mu_e\vx^2\lambda_D}{\aem\epsilon}\right)\right]^{-1},
\end{eqnarray}
where we take $n_e\sim 0.025/{\rm cm^3}$, the Coulomb logarithm $\ln\left(\mu_e\vx^2\lambda_D/\aem\epsilon\right)\sim30$ for the parameter range of interest. By demanding $\tau_{\rm cool}<\tau_{\rm acc}$, we get $\epsilon<3.4\times10^{-4}\sqrt{\mx/{\rm GeV}}$. Here we assume the epsilon-charged DM is efficiently accelerated by the Fermi mechanism. This is true when the gyroradius of the charged DM is smaller than the length of shock waves. Since the length of the shock wave can be $\sim 100~{\rm pc}$~\cite{Blandford:1978ky}, as long as $\epsilon> 5.4\times10^{-13}~(\mx/{\rm GeV})$, the charged DM will be accelerated along with baryons over a time scale $\tau_{\rm acc}\simeq10^7~{\rm years}$.

Hence, if the DM charge is in the range $5.4\times10^{-13}~(\mx/{\rm GeV})<\epsilon<3.4\times10^{-4}\sqrt{\mx/{\rm GeV}}$, the number density of the DM is negligible in the disk. We note that this constraint strongly disfavors the charged DM explanations of DAMA and CoGeNT experiments, because the experimentally preferred value is $\epsilon\sim 10^{-9}$ and $\mx\sim{\cal O}(10-20)~{\rm GeV}$. One possible way to relax this constraint is to consider the diffusion of DM into the disk. The Galactic magnetic field is not perfectly parallel, and in fact has a large nonperturbative turbulent component. The charged DM particles may diffuse into the disk as they interact with the turbulent magnetic field. The diffusion time scale can be estimated as $\tau_{\rm diff}\sim 5 H^2_{\rm d}/(3R_{\rm g}\vx)$~\cite{SanchezSalcedo:2008zd}. If the $\tau_{\rm diff}$ is smaller than the acceleration time scale $\tau_{\rm acc}$, the charged DM may be able to diffuse to the vicinity of the earth and leave a signal in direct detection experiments. This signal is sensitive to the DM number density, which highly depends on the diffusion process and will in general be smaller than $0.3~{\rm GeV/cm^3}$. We find that the charge has to satisfy $\epsilon<9\times10^{-12}~(\mx/{\rm GeV})$ with $\vx=270~{\rm km/s}$ and $B=5~{\rm \mu G}$, which is too small to fit DAMA and CoGent data.

\section{Conclusions}

We have discussed the cosmological and direct detection constraints on charged DM.  We considered in particular relic density, halo shape, large scale structure, recombination-era coupling, and direct detection constraints.  We found that charged DM must have additional annihilation modes or be non-thermally produced if it is to satisfy the CMB constraints on DM couplings to baryons, which require the DM charge be smaller than $\epsilon \sim 10^{-6}$ for 1 GeV DM (weakening to $\epsilon \sim 10^{-4}$ at 10 TeV).   We discussed the possibility that one or more of these constraints is nullified by supernova shock waves blowing charged DM out of the disk, and showed as a result that DM with epsilon charge $10^{-9}$ cannot be an explanation for the CoGeNT or DAMA excesses, though the DM-baryon interaction cross-section is large enough.  In addition, no signal in direct detection experiments could be expected in future experiments, as DM with large enough charge to generate a sizable DM-nucleus interaction cross-section would have been evacuated from the disk.

While the idea of charged DM is in many ways an elegant one, its feasibility as a DM candidate, it appears, is strongly constrained.  Further, these tight constraints also apply to any model where a massless dark photon kinetically mixes with the visible photon, so that fields charged under $U(1)_{\rm EM}$ pick up a dark charge.  This study presents constraints on a wide variety of hidden sector models that may be useful in the continued hunt for DM.

\section*{Acknowledgments}
 We thank Matt Buckley, Jonathan Feng, Manoj Kaplinghat, Aaron Pierce, Dan Feldman, and Itay Yavin for discussions.


\begin{thebibliography}{99}

\bibitem{review}
  G.~Jungman, M.~Kamionkowski, K.~Griest,
  Phys.\ Rept.\  {\bf 267}, 195-373 (1996).
  [hep-ph/9506380].

  G.~Bertone, D.~Hooper, J.~Silk,
  Phys.\ Rept.\  {\bf 405}, 279-390 (2005).
  [hep-ph/0404175].

  J.~L.~Feng,
  [arXiv:1003.0904 [astro-ph.CO]].


\bibitem{CDMS}
  Z.~Ahmed {\it et al.}  [The CDMS-II Collaboration],
  Science {\bf 327}, 1619 (2010)
  [arXiv:0912.3592 [astro-ph.CO]].

\bibitem{XENON}
  E.~Aprile {\it et al.}  [XENON100 Collaboration],
  Phys.\ Rev.\ Lett.\  {\bf 105}, 131302 (2010)
  [arXiv:1005.0380 [astro-ph.CO]].

  J.~Angle {\it et al.}  [XENON10 Collaboration],
  Phys.\ Rev.\  D {\bf 80}, 115005 (2009)
  [arXiv:0910.3698 [astro-ph.CO]].

\bibitem{champ1}
 H.~Goldberg and L.~J.~Hall,
 Phys.\ Lett.\  B {\bf 174}, 151 (1986).

\bibitem{champ2}
  A.~De Rujula, S.~L.~Glashow, U.~Sarid,
  Nucl.\ Phys.\  {\bf B333}, 173 (1990)

\bibitem{champ3}
 S.~Dimopoulos, D.~Eichler, R.~Esmailzadeh and G.~D.~Starkman,
 Phys.\ Rev.\  D {\bf 41}, 2388 (1990).

\bibitem{champ4}
 S.~Davidson and M.~E.~Peskin,
 Phys.\ Rev.\  D {\bf 49}, 2114 (1994)
 [arXiv:hep-ph/9310288].


\bibitem{champ5}
 S.~Davidson, S.~Hannestad and G.~Raffelt,
 JHEP {\bf 0005}, 003 (2000)
 [arXiv:hep-ph/0001179].

\bibitem{Chuzhoy:2008zy}
  L.~Chuzhoy and E.~W.~Kolb,
  JCAP {\bf 0907}, 014 (2009)
  [arXiv:0809.0436 [astro-ph]].

\bibitem{Dubovsky:2003yn}
  S.~L.~Dubovsky, D.~S.~Gorbunov and G.~I.~Rubtsov,
  JETP Lett.\  {\bf 79}, 1 (2004)
  [Pisma Zh.\ Eksp.\ Teor.\ Fiz.\  {\bf 79}, 3 (2004)]
  [arXiv:hep-ph/0311189].

\bibitem{Gardner:2009et}
  S.~Gardner and D.~C.~Latimer,
  Phys.\ Rev.\  D {\bf 82}, 063506 (2010)
  [arXiv:0904.1612 [hep-ph]].


\bibitem{Feng:2008mu}
  J.~L.~Feng, H.~Tu, H.~-B.~Yu,
  JCAP {\bf 0810}, 043 (2008).
  [arXiv:0808.2318 [hep-ph]].


\bibitem{Ackerman:2008gi}
  L.~Ackerman, M.~R.~Buckley, S.~M.~Carroll {\it et al.},
  Phys.\ Rev.\  {\bf D79}, 023519 (2009).
  [arXiv:0810.5126 [hep-ph]].


\bibitem{Feng:2009mn}
  J.~L.~Feng, M.~Kaplinghat, H.~Tu and H.~B.~Yu,
  JCAP {\bf 0907}, 004 (2009)
  [arXiv:0905.3039 [hep-ph]].

\bibitem{Ibarra:2008kn}
  A.~Ibarra, A.~Ringwald, C.~Weniger,
  JCAP {\bf 0901}, 003 (2009).
  [arXiv:0809.3196 [hep-ph]].


\bibitem{Kaloper:2009nc}
  N.~Kaloper, A.~Padilla,
  JCAP {\bf 0910}, 023 (2009).
  [arXiv:0904.2394 [astro-ph.CO]].

\bibitem{Dai:2009hx}
  D.~C.~Dai, K.~Freese and D.~Stojkovic,
  JCAP {\bf 0906}, 023 (2009)
  [arXiv:0904.3331 [hep-ph]].





\bibitem{dama}
  R.~Bernabei {\it et al.}  [DAMA Collaboration],
  Eur.\ Phys.\ J.\  C {\bf 56}, 333 (2008)
  [arXiv:0804.2741 [astro-ph]].

\bibitem{cogent}
  C.~E.~Aalseth {\it et al.} [ CoGeNT Collaboration ],
    [arXiv:1002.4703 [astro-ph.CO]].

\bibitem{foot}
  R.~Foot,
  arXiv:1008.0685 [hep-ph].  R.~Foot,
  Phys.\ Rev.\  {\bf D78}, 043529 (2008).
  [arXiv:0804.4518 [hep-ph]].

\bibitem{Sigurdson}
  K.~Sigurdson, M.~Doran, A.~Kurylov {\it et al.},
  Phys.\ Rev.\  {\bf D70}, 083501 (2004).
  [astro-ph/0406355].

\bibitem{Feldman}
  D.~Feldman, Z.~Liu, P.~Nath,
  Phys.\ Rev.\  {\bf D75}, 115001 (2007).
  [hep-ph/0702123 [HEP-PH]]. D.~Feldman, Z.~Liu, P.~Nath and G.~Peim,
  Phys.\ Rev.\  D {\bf 81}, 095017 (2010)
  [arXiv:1004.0649 [hep-ph]].

\bibitem{Holdom:1985ag}
  B.~Holdom,
  Phys.\ Lett.\  B {\bf 166}, 196 (1986).


\bibitem{Sommerfeld:1931}
A.~Sommerfeld,
Annalen der Physik {\bf 403}, 257 (1931).

\bibitem{Baer:1998pg}
  H.~Baer, K.~m.~Cheung and J.~F.~Gunion,
  Phys.\ Rev.\  D {\bf 59}, 075002 (1999)
  [arXiv:hep-ph/9806361].

\bibitem{Hisano:2002fk}
  J.~Hisano, S.~Matsumoto and M.~M.~Nojiri,
  Phys.\ Rev.\  D {\bf 67} (2003) 075014
  [arXiv:hep-ph/0212022].  J.~Hisano, S.~Matsumoto and M.~M.~Nojiri,
  Phys.\ Rev.\ Lett.\  {\bf 92}, 031303 (2004)
  [arXiv:hep-ph/0307216].  J.~Hisano, S.~Matsumoto, M.~M.~Nojiri and O.~Saito,
  Phys.\ Rev.\  D {\bf 71}, 063528 (2005)
  [arXiv:hep-ph/0412403].

\bibitem{sommerfeld}

  M.~Cirelli, A.~Strumia and M.~Tamburini,
  Nucl.\ Phys.\  B {\bf 787}, 152 (2007)
  [arXiv:0706.4071 [hep-ph]].
  J.~March-Russell, S.~M.~West, D.~Cumberbatch and D.~Hooper,
  JHEP {\bf 0807}, 058 (2008)
  [arXiv:0801.3440 [hep-ph]].
  M.~Cirelli, M.~Kadastik, M.~Raidal and A.~Strumia,
  Nucl.\ Phys.\  B {\bf 813}, 1 (2009)
  [arXiv:0809.2409 [hep-ph]].
  N.~Arkani-Hamed, D.~P.~Finkbeiner, T.~R.~Slatyer and N.~Weiner,
  Phys.\ Rev.\  D {\bf 79}, 015014 (2009)
  [arXiv:0810.0713 [hep-ph]].
  M.~Pospelov, A.~Ritz,
  Phys.\ Lett.\  {\bf B671}, 391-397 (2009)
  [arXiv:0810.1502 [hep-ph]].
  P.~J.~Fox and E.~Poppitz,
  Phys.\ Rev.\  D {\bf 79}, 083528 (2009)
  [arXiv:0811.0399 [hep-ph]].
  M.~Lattanzi, J.~I.~Silk,
  Phys.\ Rev.\  {\bf D79}, 083523 (2009).
  [arXiv:0812.0360 [astro-ph]].
  R.~Iengo,
 JHEP {\bf 0905}, 024 (2009)
 [arXiv:0902.0688 [hep-ph]].
 R.~Iengo,
 arXiv:0903.0317 [hep-ph]. S.~Cassel,
  arXiv:0903.5307 [hep-ph].
  T.~R.~Slatyer,
  JCAP {\bf 1002}, 028 (2010)
  [arXiv:0910.5713 [hep-ph]].
  L.~Visinelli, P.~Gondolo,
[arXiv:1007.2903 [hep-ph]].


\bibitem{freezeout1}

  J.~L.~Feng, M.~Kaplinghat and H.~B.~Yu,
  arXiv:1005.4678 [hep-ph].
  J.~B.~Dent, S.~Dutta and R.~J.~Scherrer,
  Phys.\ Lett.\  B {\bf 687}, 275 (2010)
  [arXiv:0909.4128 [astro-ph.CO]].
  J.~Zavala, M.~Vogelsberger and S.~D.~M.~White,
  Phys.\ Rev.\  D {\bf 81}, 083502 (2010)
  [arXiv:0910.5221 [astro-ph.CO]].


\bibitem{freezeout2}
  J.~Hisano, S.~Matsumoto, M.~Nagai, O.~Saito and M.~Senami,
  Phys.\ Lett.\  B {\bf 646}, 34 (2007)
  [arXiv:hep-ph/0610249].
  Q.~Yuan, X.~J.~Bi, J.~Liu, P.~F.~Yin, J.~Zhang and S.~H.~Zhu,
  JCAP {\bf 0912}, 011 (2009)
  [arXiv:0905.2736 [astro-ph.HE]].
  H.~Iminniyaz and M.~Kakizaki,
  arXiv:1008.2905 [astro-ph.CO].
  S.~Mohanty, S.~Rao and D.~P.~Roy,
  arXiv:1009.5058 [hep-ph].
   S.~Hannestad and T.~Tram,
  arXiv:1008.1511 [astro-ph.CO].
  A.~Hryczuk, R.~Iengo, P.~Ullio,
  [arXiv:1010.2172 [hep-ph]].





\bibitem{Gondolo:1990dk}
  P.~Gondolo and G.~Gelmini,
  Nucl.\ Phys.\  B {\bf 360}, 145 (1991).

\bibitem{Kolb:1990vq}
  E.~W.~Kolb and M.~S.~Turner,
  Front.\ Phys.\  {\bf 69}, 1 (1990).

\bibitem{Silk:1967kq}
  J.~Silk,
  Astrophys.\ J.\  {\bf 151}, 459 (1968).

\bibitem{Burrage:2009yz}
  C.~Burrage, J.~Jaeckel, J.~Redondo {\it et al.},
  JCAP {\bf 0911}, 002 (2009).
  [arXiv:0909.0649 [astro-ph.CO]].

\bibitem{nasa}

http://lambda.gsfc.nasa.gov/product/map/current/parameters.cfm

\bibitem{Boehm:2001hm}
  C.~Boehm, A.~Riazuelo, S.~H.~Hansen and R.~Schaeffer,
  Phys.\ Rev.\  D {\bf 66}, 083505 (2002)
  [arXiv:astro-ph/0112522].  C.~Boehm, P.~Fayet and R.~Schaeffer,
  Phys.\ Lett.\  B {\bf 518}, 8 (2001)
  [arXiv:astro-ph/0012504].  C.~Boehm and R.~Schaeffer,
  arXiv:astro-ph/0410591.




\bibitem{Lindhard:1961zz}
  J.~Lindhard and M.~Scharff,
  Phys.\ Rev.\  {\bf 124}, 128 (1961).


\bibitem{Kamionkowski:2008gj}
  M.~Kamionkowski and S.~Profumo,
  Phys.\ Rev.\ Lett.\  {\bf 101}, 261301 (2008)
  [arXiv:0810.3233 [astro-ph]].


\bibitem{Dave:2000ar}
  R.~Dave, D.~N.~Spergel, P.~J.~Steinhardt and B.~D.~Wandelt,
  Astrophys.\ J.\  {\bf 547}, 574 (2001)
  [arXiv:astro-ph/0006218].

\bibitem{Yoshida:2000bx}
  N.~Yoshida, V.~Springel, S.~D.~M.~White and G.~Tormen,
  Astrophys.\ J.\  {\bf 535}, L103 (2000)
  [arXiv:astro-ph/0002362].

\bibitem{Moore:2000fp}
  B.~Moore, S.~Gelato, A.~Jenkins, F.~R.~Pearce and V.~Quilis,
  Astrophys.\ J.\  {\bf 535}, L21 (2000)
  [arXiv:astro-ph/0002308].

\bibitem{Craig:2001xw}
  M.~W.~Craig and M.~Davis,
  arXiv:astro-ph/0106542.

\bibitem{Kochanek:2000pi}
  C.~S.~Kochanek and M.~J.~White,
  Astrophys.\ J.\  {\bf 543}, 514 (2000)
  [arXiv:astro-ph/0003483].

\bibitem{Spergel:1999mh}
  D.~N.~Spergel and P.~J.~Steinhardt,
  Phys.\ Rev.\ Lett.\  {\bf 84}, 3760 (2000)
  [arXiv:astro-ph/9909386].

\bibitem{MiraldaEscude:2000qt}
  J.~Miralda-Escude,
  arXiv:astro-ph/0002050.

\bibitem{Feng:2009hw}
  J.~L.~Feng, M.~Kaplinghat and H.~B.~Yu,
  Phys.\ Rev.\ Lett.\  {\bf 104}, 151301 (2010)
  [arXiv:0911.0422 [hep-ph]].

\bibitem{Ibe:2009mk}
  M.~Ibe and H.~B.~Yu,
  Phys.\ Lett.\  B {\bf 692}, 70 (2010)
  [arXiv:0912.5425 [hep-ph]].


\bibitem{Buote:2002wd}
  D.~A.~Buote, T.~E.~Jeltema, C.~R.~Canizares and G.~P.~Garmire,
  Astrophys.\ J.\  {\bf 577}, 183 (2002)
  [arXiv:astro-ph/0205469].

\bibitem{Humphrey:2006rv}
  P.~J.~Humphrey, D.~A.~Buote, F.~Gastaldello, L.~Zappacosta, J.~S.~Bullock, F.~Brighenti and W.~G.~Mathews,
  Astrophys.\ J.\  {\bf 646}, 899 (2006)
  [arXiv:astro-ph/0601301].

\bibitem{Markevitch:2003at}
  M.~Markevitch, A.~H.~Gonzalez, D.~Clowe {\it et al.},
  Astrophys.\ J.\  {\bf 606}, 819-824 (2004).
  [astro-ph/0309303].

\bibitem{liam}
  A.~L.~Fitzpatrick, K.~M.~Zurek,
    [arXiv:1007.5325 [hep-ph]].

\bibitem{Blandford:1978ky}
  R.~D.~Blandford, J.~P.~Ostriker,
  Astrophys.\ J.\  {\bf 221}, L29-L32 (1978).

\bibitem{SanchezSalcedo:2008zd}
  F.~J.~Sanchez-Salcedo, E.~Martinez-Gomez,
[arXiv:0812.0797 [astro-ph]]. F.~J.~Sanchez-Salcedo, E.~Martinez-Gomez, J.~Magana,
  JCAP {\bf 1002}, 031 (2010).
  [arXiv:1002.3145 [astro-ph.CO]].




\end{thebibliography}
\end{document}